\documentstyle[prl,aps,epsfig,amssymb]{revtex}
\begin{document}
\draft
\newcommand{\LS}{\langle s \rangle}
\newcommand{\INFL}{(1/\theta)}
\newcommand{\OC}{{\mathcal O}}
\newcommand{\aexpo}{b}
\newcommand{\MP}{{\mathcal P}}
\newcommand{\Nnorm}{{\mathcal N}}
\newcommand{\zunst}{\overline{z_{\text{act}}}}
\newcommand{\zst}{\overline{z_{\text{st}}}}
\newcommand{\zc}{\overline{z_{\text{c}}}}
\newcommand{\pst}{\overline{p_{\text{st}}}}
\newcommand{\pc}{\overline{p_{\text{c}}}}
\newcommand{\binomial}[2]{{#1 \choose #2}}
\newcommand{\bracketOpen}{[}
\newcommand{\bracketClose}{]}

\twocolumn[\hsize\textwidth\columnwidth\hsize\csname
@twocolumnfalse\endcsname

\title{A solvable non-conservative model of Self-Organized Criticality}
\author{Gunnar Pruessner and Henrik Jeldtoft Jensen}
\address{
Mathematics Department, 
Imperial College 180 Queen's Gate,
London SW7 2BZ,
United Kingdom\\ 
gunnar.pruessner@physics.org
and
h.jensen@ic.ac.uk 
}
\date{\today}
\maketitle
\begin{abstract}
We present the first solvable non-conservative sandpile-like {\em
 critical} model of Self-Organized Criticality (SOC), and thereby 
substantiate the suggestion by Vespignani and Zapperi 
{\bracketOpen}A. Vespignani and S. Zapperi, 
Phys. Rev. E {\bf 57},  6345  (1998){\bracketClose}
that a lack of conservation in the microscopic 
 dynamics of an SOC-model can be compensated by introducing an external
 drive and thereby re-establishing criticality. The model shown is
 critical for all values of the conservation parameter.
 The analytical derivation
 follows the lines of Br{\"o}ker and Grassberger 
{\bracketOpen}H.-M. Br{\"o}ker and P. Grassberger, 
Phys. Rev. E {\bf 56},  3944 (1997){\bracketClose} 
and is supported by numerical simulation. In the limit of vanishing
 conservation the Random Neighbor Forest Fire Model (R-FFM) is
 recovered.
\end{abstract}
]
The r\^ole of conservation in SOC-models is an old issue
\cite{Grinstein:1990,MiddletonTang:1995,Jensen:98} and is still
unsettled. 
The number of non-conservative models which
are definitely critical is, however, strikingly small. The 
Random Neighbor Forest Fire Model (R-FFM) \cite{Christensen:1993} is
one of them, while the Random Neighbor Olami-Feder-Christensen model (R-OFC)
has been shown not to be critical in the non-conservative regime
\cite{BroekerGrassberger:1997,ChabanolHakim:1997}. 
The nearest neighbor OFC model is commonly accepted to be
critical in the conservative limit, but whether this model is
critical in the non-conservative regime is still
debated\cite{MiddletonTang:1995,LisePaczuski:2001}. 

In \cite{VespignaniZapperi:1998,DickmanBraz:2000} it has been suggested that
non-conservation in the microscopic dynamics can be compensated by an
external drive in order to re-establish criticality.  Applying this 
concept directly to a model known to be non-critical in its
original definition provides the ideal basis to identify the effect of
such an external drive. 
In this letter such a model is defined and solved semi-analytically. The
results are compared to simulations and the (trivial) critical exponents
are extracted. Several limits are discussed.

The model, which is derived from the FFM \cite{DrosselSchwabl:1992} and the
Zhang model \cite{Zhang:1989}, has three main parameters: $N$ is the total
number of sites, 
which diverges in the thermodynamic limit. 
The number of randomly chosen ``neighbors'' is
given by $n$, where $n=4$ in all examples, corresponding to a
two dimensional square lattice. The conservation parameter is
$\alpha$. The degree of 
non-conservation is then $1 - n \alpha$, as it is shown below. 
Each site $i \in \{1, 2,
\cdots, N\}$ has associated a value $z_i$ for its ``energy''. Sites with
$0\le z_i<1-\alpha$ are said to be ``stable'', sites with $1-\alpha \le
z_i <1$ are called ``critical'' and sites with $1\leq z_i$ are
``active''. Negative energies are not allowed. 
The probability density function (PDF) for the variable $z_i$ is
$P(z)$ with $z \in [0,1[$ and is defined only
when no sites are active. $P(z)$ contains most of the
stationary properties of the model. 

The dynamics of the model are defined as follows: 
After an initial choice of $z_i$ ($i=1,2,. . .,N)$ from a uniform 
distribution in the interval $ [0, 1[ $, the model is updated by
repeatedly (i) ``driving'', (ii) ``triggering'' and (iii) ``relaxing'' 
the system. During the 
drive (i), $i=1,. . . ,\INFL$ sites are chosen randomly ($\INFL=p/f$ in
the notation of \cite{DrosselSchwabl:1992}), one
after the other,
and their energy $z_i$ are set to $1-\alpha$, if the site is stable,
otherwise $z_i$ remains unchanged.
Subsequently one random site $j$ is
chosen and if it is critical, the system is triggered (ii) by setting the
energy of the chosen site to $1$, i.e. making it active (initial
seed). Otherwise $z_j$ remains unchanged and the model is driven again
by repeating (i). As long as $N$ and $\INFL$ are
finite, the system will escape from the driving
loop sooner or later. In the thermodynamic limit this is ensured by a
non vanishing density of critical sites.

During the relaxation (iii) the energy of each active site $i$ is redistributed
according to the conservation parameter $\alpha$ to
$n$ randomly chosen sites $j$ and the energy of $z_i$ is then set to
$0$: 
\begin{equation} \label{eq:visit}
 z_j \to z_j + \alpha z_i \quad\quad z_i\to 0
\end{equation}
Each visit or ``toppling'' (\ref{eq:visit}) defines a microscopic
time step and dissipates exactly $(1-n\alpha)z_i$ energy units.
The sites $j$ are chosen randomly one after the other and are not
necessarily different. 
In the thermodynamic limit the probability of choosing a target site
which is already active or was already charged during the same
avalanche, vanishes and therefore the order of these visits is irrelevant.
In this very restricted sense the model might be considered as ``Abelian''. 
In contrast,
sites in finite systems have always a finite probability to get charged
more than once. Nevertheless, this probability decreases rapidly with
increasing system size. 

The number of active sites relaxed by (\ref{eq:visit}) defines the
avalanche size $s$, which is always positive due to the initial seed. In
the stationary state the avalanches must dissipate, on average,
the same amount of energy as is supplied by the external
drive and the initial seed. The average dissipation depends on the
avalanche size weighted average energy of active sites $\zunst$, which
is equivalent to the average energy of active sites per toppling.
Therefore
\begin{equation} \label{eq:average_s} 
 (1- n \alpha) \zunst \LS =
 \INFL {\pst\over\pc} (1-\alpha-\zst) + (1-\zc)
\end{equation}
must hold exactly in the stationary state even for finite systems and
does not introduce any approximation.  Here $\LS$ is the average
avalanche size, $\pst$ ($\pc$) is the 
density of stable (critical) sites (the drive stage
is, on average, repeated $1/\pc$ times), $\zst$ and $\zc$ are
the average energy of 
stable and critical sites respectively. As in
\cite{BroekerGrassberger:1997} the only crucial 
assumption is that $\LS/N$ as well as $\INFL / N$
vanishes in the thermodynamic limit, which turns out to be entirely
consistent with the results. This assumption allows us, for example,
to assume the distribution $P(z)$ to be essentially unaffected by
external drive or relaxation for sufficiently large systems. 

From (\ref{eq:average_s}) it is clear that in general
$\LS$ diverges for diverging $\INFL$ or vanishing dissipation
rate $1- n \alpha$. From the microscopic dynamics it is clear that there
is always a non-vanishing fraction of sites with $z=0$, therefore
$(1-\alpha)-\zst$ is finite and a divergence of $\INFL$ entails a
divergence of $\LS$, {\em which is a sign of criticality}.

In the following outline of the actual calculation, which is adapted
from \cite{BroekerGrassberger:1997}, the PDF's
of the model are derived.

After an avalanche, each site belongs to one of $m+2$ classes, where $m=\lfloor
1/\alpha\rfloor$. The index $k=0,1,. . .,m$ of the class indicates the
number of charges received from other toppling sites since their last
toppling, while $k=m+1$ 
indicates the class of sites, whose energy has 
been set by external drive. A site charged more than $m$
times must be active. For each of these classes a conditional
distribution function 
$Q_k(z)$ is introduced, describing the distribution of energy among
non-active sites, which have been charged $k= 0, . . .,m$ times or
externally driven, 
$k=m+1$. The distribution of sites which have not been charged since
their last toppling, $Q_0(z)$, is a delta peak at $z=0$.
For convenience the normalization of $Q_0(z)$
is chosen to be unity and all other distribution functions are
normalized relative to class $0$. The distribution of
sites, which have been driven externally and have not changed since
then, $Q_{m+1}(z)$, is obviously a delta peak at 
$1-\alpha$. If the fraction of externally driven sites is $g$, $P(z)$
can be written as
\begin{equation} \label{eq:defP}
 P(z) = \Nnorm \sum_{k=0}^{m+1} (1-g) Q_{k}(z)
\end{equation}
with appropriately chosen normalization $\Nnorm<1$. The upper bound for
the energy of an active site is the geometric sum
$1+\alpha(1+\alpha(\cdots )) = 1/(1-\alpha)$, neglecting
double charges. Therefore, the support of the distribution function
of active sites $C(z)$ is $[1, 1/(1-\alpha)[$. If this distribution is
normalized, the expected increase per avalanche in
the class $k>0$ is given, in the thermodynamic limit
(where multiple toppling can be neglected), by the convolution
\begin{equation} \label{eq:convo}
 n \LS \int_1^{1/(1-\alpha)}  \!\!\!\!\!\!\!\! 
    dz' C(z') Q_{k-1}(z-\alpha z') \ ,
\end{equation}
where the factor $n\LS$ takes into account the expected total number of
charges. There are three different ways in which the classes $k<m+1$ may
be decreased:\\
1) By charges, $Q_k(z) n \LS$\\
2) By external drive, $Q_k(z) \theta^<((1-\alpha)-z) \INFL \pc^{-1}$, 
   where
   $\theta^<$ is the Heaviside unit function with
   $\theta^<(0)=0$. \\
3) By initial seed, $Q_k(z) \theta^>(z-(1-\alpha)) /\pc$, where
   $\theta^>(0)=1$ correspondingly.\\
Adding these contributions together and assuming stationarity leads to
$m$ equations for $Q_k$, $k=1,\cdots,m$:
\begin{equation} \label{eq:Qk_from_k-1}
 Q_{k} (z) l(z) = 
\int_1^{1/(1-\alpha)}  \!\!\!\!\!\!\!\!
dz' C(z') Q_{k-1}(z-\alpha z') \ ,
\end{equation} 
where
\begin{eqnarray} \label{eq:defl}
 l(z) = 1 + \frac{\INFL}{\pc n \LS}  \theta^<((1-\alpha)-z) + \nonumber \\
\frac{1}{ \pc n \LS} \theta^>(z-(1-\alpha))
\end{eqnarray}
has been used. For diverging $\INFL$, the last term in (\ref{eq:defl})
becomes irrelevant and the RHS of (\ref{eq:average_s}) is dominated by
the first term, meaning that the initial seed becomes irrelevant
compared to the external drive.
It is reasonable to restrict the range of $\alpha$ so that single charges
cannot activate a site, $\alpha/(1-\alpha) < 1 \Leftrightarrow \alpha <
1/2$ (due to $n \alpha < 1$, this is a restriction only for $n=1$).
Then one of the Eqs. in (\ref{eq:Qk_from_k-1}) can be written as 
\begin{eqnarray}
 C \Big(\frac{z}{\alpha}\Big) = \alpha Q_{1} (z) l(z) \ ,
\end{eqnarray}
due to the particularly simple form of $Q_0(z)$.

Since 
$Q_0(z)=\delta(z)$ 
by definition and 
$\Nnorm (1-g) Q_{m+1}(z) = g \delta((1-\alpha)-z)$ 
from (\ref{eq:defP}), one further equation is necessary in order to
find $m+3$ distributions $Q_k$, $k=0,\cdots,m+1$ and $C(z)$. This
equation concerns the construction of the distribution of
active sites $C(z)$. Since active sites are created due to charging
or as the initial seed the average distribution of the number
of those sites per avalanche is given by
\begin{equation} \label{eq:defC}
 \LS C(z) = n \LS \int_1^{1/(1-\alpha)} \!\!\!\!\!\!\!\! dz' C(z') 
 P(z - \alpha z') + \delta (z-1)
\end{equation}
where the $\delta$-function represents the initial seed. 

Although it is
{\it a priori } unknown whether there exists a stable solution, or
whether it is unique, the set
of equations given above is enough to start an iteration procedure in
order to find a solution. The scalar parameters required are $\LS$ from
(\ref{eq:average_s}), $\pst$, $\pc$, $\zst$, $\zc$, which are easily
derived from (\ref{eq:defP}) and $\zunst$, the first moment of
$C(z)$. While $n$ and $\INFL$ parameterize the problem, $g$ remains the
only unknown quantity, which is found to be
\begin{equation} \label{eq:defg}
 g = \frac{\pst \INFL}{n\LS \pc + 1}
\end{equation}
by comparing the in- and outflow of class $m+1$, the externally driven
sites, per avalanche. 

All the equations above can alternatively be derived directly from the
fundamental dynamics. This ensures that the solution is exact 
in the thermodynamic limit given the stationarity assumption.
 
The implementation of the iteration procedure is straight-forward. As
a criterion for termination, one could check whether $C(z)$, as defined by
(\ref{eq:defC}), is properly normalized \cite{BroekerGrassberger:1997},
as it can be proven that it must be correctly normalized if it is a
solution. However, it would be sufficient to assume $C(z)$
proportional to the RHS in (\ref{eq:defC}). Moreover, in the numerical
procedure 
the quality of the normalization of $C(z)$ depends strongly on the
resolution of the grid chosen, whenever $C(z)$ changes rapidly
as function of $z$. Therefore convergence of the
iteration procedure is better verified by checking whether
$C(z)$ approaches a fixed point, i.e. is invariant under
(\ref{eq:defC}). Since the distribution is expected 
to be highly non-analytic - there are at least two $\delta$-functions in
$P(z)$ - sophisticated integration routines are inappropriate.
For $n=4$ the procedure quickly converges for $0.07 < \alpha < 0.24 $, 
all non-pathological initial
values tested lead to the same stable solutions. Only for small values
of $\LS$, 
when the $\delta$-peak of the initial seed starts to propagate through
the distribution, a large grid is required for sufficient resolution.
The same problem appears close to the commensurable
limits mentioned below.
\begin{figure}[t]
\begin{center}
      \includegraphics[width=1.0\linewidth]{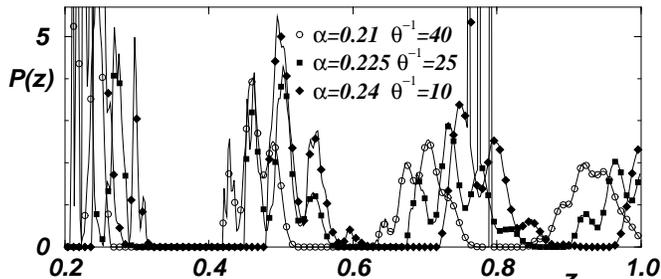}
\caption{\label{fig:simu_versus_calc} Distribution of energy, $P(z)$,
 for $n=4$ and different values of $\alpha$. Since there is only a
 $\delta$-peak in $0\le z \le 0.2$, results 
 for $z<0.2$ are cut off. Continuous lines indicate results from theory
 (grid size $32000$, integrated in $125$ bins), points 
 represent results from numerical simulations with $N=10^6$,
 $10^6$ avalanches for equilibration and $5\ 10^6$ avalanches for
 statistics ($125$ bins for the histogram).}
  \end{center}
\end{figure}
In Fig.~\ref{fig:simu_versus_calc} numerical 
simulations of the model are compared to the numerical solution of the
analytical approach. Although the PDF is very structured,
discrepancies are small and can be reduced by increasing the
resolution of the underlying grid.

\begin{figure}[t]
\begin{center}
      \includegraphics[width=1.0\linewidth]{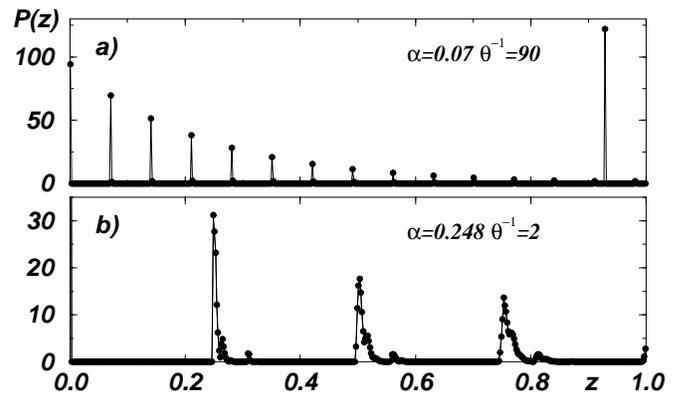}
\caption{\label{fig:alpha0} $P(z)$ as in
 Fig.~\ref{fig:simu_versus_calc}, but for $\alpha=0.07$ and
 $\alpha=0.248$ and $500$ bins.}
  \end{center}
\end{figure}

The distribution $C(z)$ collapses to a $\delta$-function in at least two
limits. 
Firstly, when $\alpha \to 0$, the FFM-limit, the probability for a site to 
become critical due to a number of charges vanishes as
$q^{(1-\alpha)/\alpha}$, where $q<1$ is the product of the probability that
a site receives a charge from a relaxing site and the probability that
a site is not driven externally between two hits. 
Hence, for $\alpha\to 0$ the mechanism of ``growing by charges''
becomes negligible and the external drive becomes the dominating source
for critical sites. 
The dynamics now become equivalent to the FFM: 
stable sites = empty sites, critical sites = trees, and active sites = fires.
Furthermore, 
as $\alpha \to 0$, the support of $C(z)$ becomes
smaller and smaller and the distribution of active sites is strongly
peaked at $z=1$, collapsing to a $\delta$-function. Therefore, the
distributions $Q_k$, $k>0$ are less smeared 
out, as shown in Fig.~\ref{fig:alpha0}(a) for a small value of
$\alpha$. Assuming $C(z)$ to be a 
$\delta$-function, one can easily reconcile the results in
\cite{Christensen:1993} (Eqs. (3) and (7)).
The assumption that $P(z)$ is unaffected by single avalanches
corresponds to $p,f \to 0$ in
the SOC-limit of the FFM \cite{ClarDrosselSchwabl:1996}.

In the second limit,
$\alpha \to 1/n$, the model becomes conservative, but more important,
$\alpha$ becomes more and more commensurable in the sense that a site
charged $m = n$ times is almost always active and therefore the support
of $Q_m$ vanishes, as it is squeezed between $n\alpha$ and $1$. 
When $\alpha$ is very close to $1/n$, most of the
active sites are provided by $Q_{m-1}$ and their average energy is just
above $1$, i.e. $C(z)$ becomes more and more $\delta$
shaped and so do the $Q_i$, as shown in Fig.~\ref{fig:alpha0}(b). The
same behavior is obtained whenever $k \alpha = 1 $ for $k \in
\mathbb{N}$.

The critical exponents of the model, 
can be obtained by mapping it on to a
branching process \cite{Harris:1963} in order to identify the
critical exponent $\aexpo=2$, where $\MP(t) \propto t^{-\aexpo}$ 
is the exponent
of avalanche duration. The exponent $\tau=3/2$, found by mapping the
model on to a random walker along an absorbing barrier, is the
exponent of avalanche sizes, $\MP(s) \propto
s^{-\tau}$. Formally these exponents arise only for diverging cutoffs in
the distributions, which are controlled by the average number of active
sites produced per single toppling, the branching ratio $\sigma$.
The cutoffs diverge for $\sigma \to 1$. 

However, the mapping is non-trivial, except when $C(z)$ is 
a $\delta$-function. This is because a distribution of active sites entails a
distribution in the branching ratio, i.e. the branching probability
itself becomes a random variable. This problem is removed by considering
the ensemble average with an annealed rather than quenched disorder in
the branching probability, i.e. writing the probabilities for one given node
(active site) to branch into $k$ new nodes (active sites) as
\begin{equation}
 \MP (1 \to k) = \Big\langle \binomial{n}{k} p^k (1-p)^{n-k} \Big\rangle_p
\end{equation}
where $p$ denotes the branching probability (which is a function of the
energy of the site) and $\langle \rangle_p$ denotes the weighted average
over the probabilities. Therefore $\sigma = \sum_k k \MP (1 \to k)$. 
This branching process is characterized by the
same generating functions as the standard branching process
\cite{Harris:1963}, which becomes critical for $\sigma = 1$.
Hence the condition for criticality is 
\begin{equation}
\sum_{k=0}^\infty k 
\Big\langle \binomial{n}{k} p^k (1-p)^{n-k} \Big\rangle_p = 
n \langle p \rangle_p = 1
\end{equation}
which is the (average) branching ratio, according to (\ref{eq:defC})
given by 
\begin{equation}
  n \langle p \rangle_p = 
n \int_1^{1/(1-\alpha)}\!\!\!\!\!\!\!\!\!\! dz
\int_1^{1/(1-\alpha)}\!\!\!\!\!\!\!\!\!\!\!\! dz' \ 
C(z') P(z - \alpha z') = 1 - \frac{1}{\LS}\ .
\end{equation}

Moreover $C(z)$ is time or generation dependent, since it evolves
from an initial distribution which is only a
$\delta$-function at $z=1$. The deviation of $C(z)$ from the final
distribution decays exponentially fast, which can be seen by
investigating the Markov chain of the repetitive convolution of a now
time dependent $C(z)$ with $P(z)$ as in (\ref{eq:defC}). Therefore the
cutoff, introduced by the deviation 
of $\sigma$ from $1$, is dominated by the asymptotic iteratively
stable limit of $C(z)$ only. Since the asymptote is approached
exponentially fast 
the transient cannot influence the value of the exponents.  The same arguments 
apply for the random walker approach, therefore $\aexpo=2$ and $\tau=3/2$ is 
true for all $\alpha \in ]0, 1/n [$.

The calculations above are {\it a priori} valid only in the
thermodynamic limit. However, a simulation of the model must consider
a finite system. Moreover the model relies on several assumptions, which
entail certain finite size scaling: $\INFL_N / N $ (the index indicates
the value to be measured in a system of size $N$) as well as $\LS_N / N $
must vanish for diverging $N$, while $\INFL_N / \LS_N$ must remain
constant. It is a well known problem in the FFM
\cite{ClarDrosselSchwabl:1994} that the 
number of trees grown between two ignitions is a parameter, $\INFL_N$,
which needs to be {\it tuned} according to the system size; it is
supposed to diverge, but its value is restricted by system size. An
inappropriatly chosen parameter produces a small value of the cutoff
or a bump in the 
distribution  function of avalanche sizes. Nevertheless, $P(z)$ depends
only weakly on $\INFL_N$. As a more 
quantitative measure for the  ``right choice of $\INFL_N$'' 
we compared $g_N$ to $g$ (see Eq. (\ref{eq:defg})) in the thermodynamic limit.
Assuming a cutoff of order $\OC(N)$ in $\MP(s)$ of a finite system, the
scaling is $\LS_N = \int ds\MP(s) s \in \OC(N^{1/2})$ and thus
$\INFL_N \in \OC(N^{1/2})$. For a more quantitative 
picture one can map the avalanche on to a random walker along an absorbing
barrier with time dependent walking probability $p(t)$ (in the sense of
\cite{Fisher:84} a drinking rather than a drunken random
walker). However, the result is comparatively involved and gives only
rough estimates for the avalanche size as a function of $N$ and $\INFL$.

In summary, a solvable SOC model, critical in the entire regime
of the conservation parameter, has been defined and the main properties
have been derived. The critical exponents are as expected the trivial
exponents of a critical branching process and a random walker. The model
clarifies the r{\^o}le of the external drive and represents an explicit
example of the recovery of criticality by introducing an external drive.

The authors acknowledge the support of EPSRC. G.P. wishes to thank
Yang Chen and Richard Rittel for useful discussions as well as Quincy
Thoeren for hospitality.
\bibliography{articles,books}
\end{document}